\documentstyle[12pt]{article}                                                                                                           
\begin{document}
\begin{center}
{\large\bf An Implementation of the Polynomial Lie Algebra Methods
for Solving a Class of Nonlinear Models in Quantum Optics
\footnote{The work was supported by the Russian Foundation for Basic
Research (grants No.00-02-81023 Bel 2000-a and No.00-02-16337).} } \\
[0.1cm] Valery P. Karassiov$^{1,2}$, Alexander A. Gusev$^{3}$,
Sergue I. Vinitsky$^2$
\\ [0.1cm]
$^1$ P.N. Lebedev Physical Institute, Leninsky Pr. 53, Moscow, 117924
Russia\\
$^2$ BTPL, JINR, Dubna, Moscow region, Russia\\
$^3$ SCAI, JINR, Dubna, Moscow region, Russia
\end{center}
\begin{abstract}
We develop some calculation schemes to determine dynamics of
a wide class of integrable quantum-optical models using their symmetry
adapted reformulation in terms of polynomial Lie algebras
$su_{pd}(2)$. These schemes, based on  "diagonal" representations
of model evolution operators (via diagonalizing Hamiltonians
with the help of the $su_{pd}(2)$ defining relations), are implemented
in the form adapted for numerical calculations. Their  efficiency is
demonstrated on the example of the second-harmonic-generation model.
\end{abstract}
\section{Introduction}
During last decades a great attention is being paid to examine
different quantum-optical models with  Hamiltonians given by
nonlinear functions in Lie algebra generators since they enable
to reveal new physical effects and phenomena (see, e.g., [1-10]
and references therein). To analyze such models one uses mainly
numerical calculations [10] because standard Lie-algebraic
techniques well adapted for solving problems with linear (in Lie
algebra generators) Hamiltonians [11], are non-efficient here, and
most of other analytical techniques (e.g., the algebraic Bethe
ansatz [12]) require in general cases tedious calculations and do
not yield simple analytical expressions for physical quantities.
On other hand, standard numerical calculation schemes dealing
with initial formulations of models [10] are limited by computer
powers and are not adapted to reveal many peculiarities of model
dynamics [8,9].

However, recently  a new universal Lie-algebraic approach,
essentially improving both analytical and numerical solutions of
physical problems, has been suggested in [4,5] and developed in
[5-9] for some nonlinear quantum models whose Hamiltonians $H$
have invariance groups $G_i:\, [G_i, H]=0$. It is based on
reformulations of models under study in terms of (introduced in
[4,5,13]) polynomial Lie algebras (PLA) $g_{pd}$ as dynamic
symmetry algebras $g^D: g^{D}=g_{pd}$ completely describing model
dynamics. Specifically, two analytical approximations of
evolution operators $U_H(t)$ were found in [8]  for a wide class
of quantum-optical models with $g^{D} = g_{pd}=su_{pd}(2)$.
Furthermore, appropriate Lie-algebraic path-integral schemes for
solving physical problems were developed in [9]. They enable to
examine model dynamics at quasiclassical levels described by
"large" values of certain characteristic parameters [8]. However,
up to now systematic examinations of their accuracy and
efficiency were absent because $su_{pd} (2)$-techniques of
getting exact solutions [5-7] yield only difference and
difference-differential equations  and  fairly complicated  (for
practical calculations) algorithms rather  than closed analytical
expressions required for this aim.
 In the present work we cancel in part this lacune. Specifically,  after 
some preliminaries (Section 2) we develop an algorithm and routines to
implement an exact calculation scheme for determining diagonal
representations of model evolution operators via solving a
spectral problem by means of the $su_{pd}(2)$-techniques [5]
(Section 3). Their efficiency is demonstrated with the help of the
computer experiments for the second-harmonic-generation model
which are also used to investigate the accuracy of the
appropriate quasiclassical approximation obtained in [8] (Section 4).

\section{Models and their symmetry adapted $su_{pd}(2)$ - formulations}
We consider a wide-spread class of quantum-optical models with
Hamiltonians [1,3]
$$ H^{mps}(m;n) = \hbar\left[\sum _{i=1}^m \omega_i a_i^+ a_i  +
\omega_0 a_0^+ a_0 + g a^+_{i_1}\dots a^+_{i_n} a_0 +
g^*a_{i_1}\dots a_{i_n} a^+_0\right],$$
$$ 1\leq i_1\leq i_2\dots\leq i_n\leq m, \quad n\geq 2    \eqno (2.1a)$$
$$ H^{D} = \hbar\left\{\omega_1 a_1^+ a_1 + \sum _{i=1}^M \left [\sigma_0
(i)\epsilon/2\,+ \,g \sigma_{+} (i)(a _1)^n + g^* \sigma_{-} (i)(a _1^+)^n
\right ]\right\}  \eqno (2.1b)$$
where  $a_{i}, a _{i}^+$ are  operators of field modes with
frequencies $\omega_i$, $\sigma_{\alpha}(i)$ are Pauli matrices,
$\epsilon$ is an energy difference of two-level atoms, $g$ are
coupling constants, and non-quadratic parts of Eqs. (2.1a) describe
different  processes of multiphoton scattering including non-degenerated
frequency conversion ($H^{mps}(n;n)$) and higher harmonics generation
($H^{mps}(1;n)$) while Eq. (2.1b) yields $n$-photon point-like
Dicke models of matter-radiation interactions [1,8]. Hilbert spaces
$L(H^{mps})$ are Fock spaces $L_F(k\leq m)=Span\{|\{n_i\}\rangle\propto
\prod_{i=0}^k (a_i^+)^{n_i}|0\rangle\}$  and $L(H^{D})=L_F(1)\otimes L_a,
\,L_a=Span\{\prod_{i=1}^M|\pm\rangle(i)\}$ where $|\pm\rangle(i)\}$ are
eigenstates of $i$-th atom [3,4].

According to [5,9] Hamiltonians (2.1) have invariance groups
$$G_i=C_n\otimes\prod_j U^j(1),\, C_n=
\{a^+_{i_\alpha} \rightarrow \exp(i2\pi k /n) a^+_{i_\alpha}\}, U^j(1)=
\{\exp(i\beta_j R_j\}   \eqno (2.2)$$
where for $H^{mps}(m;n) \,R_j\in Span\{a^+_ia_i=N_i\}$  and for
$H^{D}\,R_j$ are functions in $\sum_{i=1}^M \sigma_{\alpha} (i)$ and, besides, 
$G_i$ has
the permutation group $S_M=\{\sigma_{\alpha}(i) \rightarrow
\sigma_{\alpha}(j)\}$ as an extra factor.
Therefore, one can introduce (via the $G_i$-invariant polynomial
Jordan mappings [5,9]) two sets of collective  operators:
integrals of motion $R_j= R_j(a_i^+, a_i, \dots), j=1,\dots$  and
$G_i$-invariant dynamic variables $V_{\alpha=0,\pm}=V_{\alpha=0,\pm}
(a_i, a_j^+,\dots)$ obeying the commutation relations
$$[V_0, V_{\pm}]= \pm V_{\pm}, \; [V_-, V_+] = \phi (V_0;\{R_j\} )\equiv
\psi (V_0+1;\{R_j\}) -\psi (V_0;\{R_j\}),  $$
$$\psi (V_0;\{R_j\})=A_{\psi}\prod_{i=1}^{n_{\psi}} (V_0 +
B^{\psi}_i(\{R_j\})), \quad n_{\psi}\geq 3   \eqno (2.3a)$$
which resemble those for the $su(2)$ generators $Y_{\alpha}$ but with
polynomial structural functions $\psi (V_0;\{R_j\})$ depending
additionally
on invariant operators $R_i: \, [V_{\alpha}, R_j] = 0$
unlike the quadratic function
$$\psi (Y_0; J)=  (Y_0 + J)\; (J - Y_0 + 1)   \eqno (2.3b)$$
for $su(2)$ [9]. Therefore, $V_{\alpha}$ can be considered as generators
of PLA $su_{pd}(2)$ acting on $L(H)$ complementarily to $G_i$ [5], i.e.,
$[G_i,su_{pd}(2)]=0$ and $L(H)$ are decomposed into direct sums
$$L(H)\! =\!\sum_{[l_i]}\! L([l_i]), \! L([l_i])=\mbox{Span}\{|[l_{i=0,1,
 \dots}];f\rangle\!=\!{\cal N}(f;[l_i]) V_+^f|[l_i]\rangle\},  \! V_0
 |[l_i];f\rangle \! = \! (l_0+f) |[l_i];f\rangle,$$
$$ R_i |[l_i];f\rangle = l_i |[l_i];f\rangle,\;  N(f;[l_i])=
\prod_{r=0}^{f-1}[\psi(l_0+f-r)]^{-1/2}, \; N(0;[l_i])=1,
 \; V_-\,|[l_i]\rangle = 0,\eqno (2.4a)$$
 $$ \langle [l_i];f |[l'_i];f'\rangle =
\delta_{[l_i][l_i']}\,\delta_{ff'},
 \qquad
 I=\sum_{[l_i]}I_{[l_i]}=\sum_{[l_i],f}|[l_i];f\rangle\langle[l_i];f|
 \eqno (2.4b)$$
of $G_i\otimes su_{pd}(2)$-irreducible subspaces $L([l_i])$ with  finite
dimensions $d([l_i])$ where $|[l_i]\rangle\in L(H)$ are  "lowest"
weight basic vectors, quantum numbers $l_i, i=0,1, \dots$ are eigenvalues
of operators $R_i$ and "lowest" weight operators $R_0$ are determined by
the $su_{pd}(2)$ Casimir operators
$$\Psi (R_0)\equiv \psi (V_0;\{R_j\})-V_+ V_- = \psi (V_0+1;\{R_j\})-V_-
V_+ , \quad [V_{\alpha},\Psi (R_0)]=0 \eqno (2.4c)$$
which, by construction  (due to the invariant  theory [5]), satisfy the
characteristic identities
$$ \Psi (R_0)|_{L(H)}\equiv\, 0\, \Rightarrow \Psi (l_0) =
  \psi (l_0;\{l_j\})  \,=\, 0\,=\,\psi (l_0 + d([l_i]);\{l_j\})
= \Psi (l_0+ d([l_i]))    \eqno (2.4d)$$
which are useful in applications [5-9].

In the general cases the $G_i$-invariant Jordan mappings can be given as
follows [9]:
$$V_+ =  a^+_{i_1}\dots a^+_{i_n} a_0, \quad V_- = a_{i_1}\dots
a_{i_n}a_0^+,
\quad V_{0} = (\sum\nolimits_{i=1}^m N_i - N_0) / (n+1),$$
$$R_{1} = (\sum\nolimits_{i=1}^m N_i +n N_0) / (n+1),\qquad R_{1<j\leq m}
= \mu_{j}  N_{j-1} -  \mu_{j-1} N_j
\eqno (2.5a)$$
for models (2.1a) ($ \mu_{j}$ is the multiplicity of the $a^+_{j}$
appearancein $V^{+}_{i_1\dots\i_n}$) and
$$ V_{0} = J_0,\; V_{+} =J_{+}\,(a_1)^{n},\;
V_{-} = J_{-} (a^+_1)^n,\; J_{0}\,=\,\frac{1}{2}\sum_{i=1}^M\,
\sigma_0 (i), \; J_{\pm}\,=\,\sum_{i=1}^M\,\sigma_{\pm} (i),$$
$$R_{1} = J_0 + ({a^+_1 a_1})/{n},\qquad R_2 = J,\;
J (J + 1) = J_0^2 +  ( J_+ J_- +  J_-  J_+)/2
\eqno (2.5b)$$
for the model (2.1b).

Appropriate structural polynomials (2.3a) are determined from Eqs. (2.5),
(2.4c)-(2.4d). Specifically, they are given by expressions [9]
$$\psi_D ( V_{0};R_1,R_2)=(R_2 +  V_{0}) (R_2 +
1 -  V_{0}) (n R_1 + n - n  V_{0})^{(n)}, \eqno (2.6a)$$
$$\psi_{hg} (V_{0};R_1)=( R_1 + n V_{0})^{(n)} ( R_1 + 1 -
 V_{0}), \eqno (2.6b)$$
for the model (2.1b) and spread specifications of the model (2.1a) with
$H^{mps}(1;n) \equiv H^{hg}(n)$ (high-harmonics-generation models)
respectively; here $(A)^{(m)}\equiv A(A-1)\dots (A-m+1)$. In the general
case for the model (2.1a) we get fairly complex expressions (see, e.g.,
them for $H^{mps}(n;n)\equiv H^{fc}(n)$ in [9]), and, therefore, a special
(based on the Symbolic Computer Algebra System [14]) routine was
developed to get such expressions automatically.

The introduction of the collective operators above enables us to
express  all Hamiltonians (2.1)  as follows [5,9]:
$$ H = \hbar [\Delta V_0 + g V_+ + g^* V_- + C(\{R_i\})],\quad
[V_{\alpha},
C] = 0,   \eqno (2.7)$$
where coefficients $\Delta $ and functions $C=C(\{R_i\})$ are
found from Eq. (2.1) with the help of Eqs. (2.5) for $R_i, V_{\alpha}$.
In such a manner we get [9]:
$$C^{hg}(R_1)\,=\,(\omega_1\,+\,\omega_0) R_1 ,\qquad \Delta^{hg}\; =\; n
\omega_1\,-\, \omega_0 \quad \mbox {for}\; H^{hg}(n), \eqno (2.8a)$$
$$C^{fc}([R_i]) = \omega_0 R_1 +  \sum\limits_{j=1}^{n}\frac{\omega_j}{n}
\left(R_1 - \sum\limits_{i=1}^{n-1} i R_{i+1}\right) +
\sum\limits_{j=1}^{n-1}\omega_j \sum\limits_{i=j+1}^n\, R_i,\;\Delta^{fc}
= \sum_{i=1}^n\omega_i -\omega_0,  \eqno (2.8b)$$
$$C^{D}(R_1,R_2)\, =\, n\,R_1\, \omega_1,\quad \Delta^{D} \;=\;\epsilon -
\,\omega_1  \eqno (2.8c)$$
for $ H^{hg}(n), \; H^{fc}(n)\equiv H^{mps}(n;n)$ and $H^{D}$
respectively. At the same time decompositions (2.3) of $L(H)$ are specified
by determining quantum numbers $l_i$ (integral of motions) and "lowest"
weight vectors $|[l_i]\rangle\in L(H)$ as solutions of defining relations in
(2.4a) [4,5]. Specifically, in such a manner one gets [9]:
$$L^{hg}([l_i])\equiv\! L^{hf}\!\left(l_0=\frac{\kappa-s}{1+n}, l_1=
\frac{\kappa +ns}{1+n}\right), \, |[l_i(\kappa,s)]\rangle=
\frac{(a^+_1)^{\kappa}(a^+_0)^s}{[\kappa!
s!]^{1/2}}|0\rangle,\,0\leq\kappa
\geq n-1,s\geq 0,  \eqno (2.9a) $$
$$L^{fc}([l_i])\equiv L^{fc}\left[l_0=(\sum\limits_{i=1}^n\kappa_i-s)/
(1+n),l_1= (\sum\limits_{i=1}^n \kappa_i +ns)/(1+n), l_{2\leq i
\leq n} = \kappa_{i-1} - \kappa_{i}\right],$$
$$|[l_i({\kappa_i},s)]\rangle =\frac{\prod_{i=1}^n (a^+_i)^{\kappa_i}
(a^+_0)^s}{[\prod_i\kappa_i! s!]^{1/2}}|0\rangle,\quad\kappa_i\geq 0,s\geq
0, \;\prod_{i=1}^n \kappa_i = 0,  \eqno (2.9b)$$
$$L^{D}([l_i])\equiv L^{D}(l_0=-j = -l_2, l_1= -j +\frac{\kappa}{n};
\{l_{r\geq {2+r}}=j_r\}),\quad\kappa\geq 0,j=\frac N2-\left[\frac
N2\right],
\dots,\frac N2,$$
$$|[l_{i\leq 2}(\kappa,j)];\{j_r\}\rangle =|\kappa\rangle_f\otimes
|j;\{j_r\};-j\rangle_a,\, |\kappa\rangle_f=[\kappa!]^{-1/2}
(a^+_1)^{\kappa}|0\rangle \in  L_F(1) \eqno (2.9c)$$
where $|j;\{j_r\};-j\rangle_a$ are "lowest" weight vectors of the
$su(2)$-irreducible "atomic" subspaces $L(j;\{j_r\})\subset L_a$ and extra
integral of motions ("intermediate atomic quasispins"
[3,9]) $j_r$ are due to the factor $S_M$ in $G_i^D$; in all these cases
$d([l_i])=s+1$, where  $s= \min\{2j,\kappa/n\}$ for $H^D$.

So, Eqs. (2.4)-(2.9) yield the "$su_{pd}(2)$-cluster" formulation of
models (2.1) entailing a complete description of model dynamics with 
the help of
the $su_{pd}(2)$ algebra formalism.  In particular, it provides adequate
representations [5-9] for model evolution operators $U_H(t)$ to calculate
quantum expectations $\langle O(t) \rangle$ of any operators $O=O(a^+_i,
\dots)$ acting on $L(H)$:
$$\langle O(t) \rangle=
\mbox{Tr} [U_{H}(t) \,\rho \,U^{\dagger}_{H}(t)\,O]
\,=\,\sum_{[l_i]}\sum_{f}
\langle [l_i];f|\,\rho \,U^{\dagger}_H (t) O U_H (t)\,|[l_i];f\rangle.
\eqno (2.10)$$
\section{Diagonal representations of model evolution operators and
algebraic schemes to determine them}

One of adequate representations of $U_H(t)$  has a diagonal form [8,9]
$$U_H(t)\; =\; \sum_{[l_i];v} \,e ^{-it\epsilon ([l_i];v)}\;
|E_v([l_i]) \rangle\,\langle E_v([l_i])|, \quad E_v([l_i])=
\hbar \epsilon ([l_i];v)= \hbar [C(\{l_i\}) + \lambda_v([l_i])] \eqno
(3.1)$$
where $E_v([l_i])$ and $\{|E_v([l_i])\rangle\}$ are, respectively,
eigenvalues and complete sets of  orthonormalized
eigenvectors of the Hamiltonian (2.7):
$$H |E_f([l_i])\rangle = E_f([l_i]) |E_f([l_i])\rangle,\;
\langle E_f( \delta_{ff'},\; I=\sum_{[l_i],v} |E_{v}([l_i])\rangle
\langle E_{v}([l_i])|  \eqno(3.2)$$ In the "linear" case, when
$n_{\psi} = 2 $ in (2.3a), $\psi (V_0;\{R_j\}) =\psi (Y_0; J)$
and PLA $su_{pd}(2)$ are reduced to the familiar $su(2)$ algebra,
the eigenproblem (3.2) is solved exactly [8] with the help of the
$SU(2)$ displacement operators $ S_{Y}(\xi=r{g}/{|g|})= \exp(\xi
Y_+ - \xi^* Y_-)$ [11]; herewith solutions are given by simple
analytical expressions [5-8].

However, it is not the case when $n_{\psi} \geq 3 $ in (2.3a) in
view of the absence of explicit expressions for matrix elements
$\langle [l_i]; f| \exp(\sum_i a_i V_i) |[l_i]; v\rangle$ [8]. In
fact, in [6] an algorithm has been developed to solve
eigenproblem (3.2) with $n_{\psi}\geq 3$ via "dressing"  an
auxiliary "linear" (with $n_{\psi}=2$ one; but it is unsuitable
for practical calculations, and, really nowadays it is known only
an approximate analytical solution [8] of the problem  (3.2)
given by  approximate $SU(2)$-quasiclassical eigenfunctions
$$|E^{qc}_v([l_i];\xi)\rangle\equiv
|E^{SU(2)}_v([l_i];\xi=r{g}/{|g|})\rangle
= \exp(\xi Y_+ -\xi^* Y_-) |[l_i];v\rangle = \sum_f S^j_{f v}(\xi)|[l_i];
f\rangle,$$
$$S^j_{f v}(\xi)=({g}/{|g|})^{f-v}d^{j}_{-j +f,-j +v}(2r),\quad 2j= s=
d[l_i]-1 \eqno(3.3a)$$
and eigenenergies
$$E^{qc}_v([l_i];\xi)=\langle E^{qc}_v([l_i];\xi)| H |E^{qc}_v([l_i];
\xi) \rangle=\hbar [C(\{l_i\})+ \lambda_v^{qc}([l_i];r)],
\; \lambda_v^{qc} ([l_i];r)= \Delta (j +l_0)$$
$$ -\Delta (j - v)\cos 2r +2|g| \sum_{f=0}^s\,\sqrt{(s-f)(f+1)\phi (-j+f)}
\,d^{j }_{-j+f, -j+v} (2r)\,d^{j }_{-j +f+1, -j +v} (2r) $$
$$\approx\lambda_v^{cmf}([l_i];r) = \Delta (j +l_0) -\Delta (j - v)
\cos 2r-2|g|\,(j - v)\, \sin 2r\,\sqrt{\phi [(- j+v)\cos 2r]},$$
$$\phi (- j+f)\,\equiv \,\langle [l_i];f|{\psi (V_0+1;\{R_i\})}/
{\psi (Y_0+1;j)}|[l_i];f\rangle \eqno(3.3b)$$
where generators $Y_{\alpha}$  of the $su(2)$ algebra are  connected with
those of PLA $su_{pd}(2)$ via generalized Holstein-Primakoff mappings
[5,8], $d^{j}_{m,n} (2r)$ is  the $SU(2)\, d$- function expressed in terms 
of the Gauss hypergeometric function [15] and approximate values 
$\lambda_v^{cmf}
([l_i];r)$ are calculated in the cluster mean-field approximation:
$\langle [l_i];f|F (V_{\alpha})|[l_i];f\rangle = F (\langle [l_i];f|
V_{\alpha}|[l_i];f\rangle)$.  Values of the parameter $r$ in (3.3b) are found
from energy-stationarity-conditions and/or from minimizing a proximity
measure of Hamiltonians $H$ and $H^{qc}(\xi)=\sum_{v,[l_i]}\,
E^{qc}_v([l_i];\xi) |E^{qc}_v([l_i];\xi)\rangle \langle E^{qc}_v([l_i];\xi)|$; 
a standard measure for such estimates on the subspaces $L([l_i])$
is  defined with the help of the euclidean operator norm as follows  [5,8]
$$\delta^2_H ([l_i])=\frac{Tr_{[l_i]}(H - H^{qc}(\xi))^2} {Tr_{[l_i]} (H -
C(\{l_i\}))^2}=\frac{\sum_{v}\left[(\lambda_v ([l_i]))^2 -(\lambda_v^{qc}
([l_i];r))^2\right]}{\sum_{v}(\lambda_v ([l_i]))^2}.
\eqno (3.4a)$$

However, according to the general quasiclassicality theory [16] all
approximations (3.3) are valid only for large values of $d([l_i])$, and,
besides, the measure (3.4a) gives only a global rather than local
characteristic of the approximate energy spectra $\{E^{qc}_v([l_i];\xi)\}$
that does not allow to feel their important symmetry properties and local
peculiarities related to "energy errors"
$$\Delta E_v([l_i]) = \hbar\,
[\lambda_v([l_i])\,-\,\lambda_v^{qc}([l_i];r)]
\equiv \delta E_v([l_i]) \cdot E_v([l_i])
\eqno (3.4b)$$
Therefore, it is useful to have  convenient  algorithms to get exact
solutions of the eigenproblem (3.2) using $su_{pd}(2)$ defining relations.

Specifically, taking into account  Eqs.  (2.4a), one can look for
eigenfunctions $|E_f([l_i])\rangle$ on each $su_{pd}(2)$-irreducible
space $L([l_i])$ in the form [5,8]
$$|E_f([l_i])\rangle = \sum_{v=0}^s Q_v^f([l_i]) |[l_i];v\rangle
 =\sum_{v=0}^s \,\tilde Q_v^f([l_i])\, V_{+}^v |[l_i]\rangle,\quad
 \tilde Q_v^f([l_i])=N(v;[l_i])\, Q_v^f([l_i])
\eqno(3.5a) $$
where $s=d[l_i]-1$ and, in view of Eqs. (3.2) amplitudes $Q_v^f([l_i])$
satisfy the following orthonormalization and completeness conditions:
$$ \sum_{v=0}^s Q_v^f([l_i])\; Q_v^{f'}([l_i]) \,=\, \delta_{ff'},
\qquad \sum_{f}^s Q_v^f([l_i])\;Q_{v'}^f([l_i])\, =\, \delta_{vv'}
\eqno(3.5b) $$
Then, inserting Eq. (3.5a) for $|E_f([l_i])\rangle$ and Eq. (2.7) for $H$
in the first equation of (3.2) and using Eqs. (2.3a), (2.4c)-(2.4d), one gets
a set of recurrence relations
$$ g^{*}\psi (l_0+v+1; [l_i])\tilde Q_{v+1}^{f}([l_i]) =
[\lambda_f([l_i])-
\Delta (v + l_0)]\tilde Q_v^{f}([l_i])) - g\tilde Q_{v-1}^{f}([l_i]), \,
v,f=0,\cdots,s    \eqno (3.6)$$
which together with boundary conditions $\tilde Q_{-1}^{f}([l_i]) = 0 =
\tilde Q_{s+1}^{f}([l_i])$ determine amplitudes $Q_v^{f}([l_i])$ and
eigenenergies $E_{f}([l_i])$ from solutions of the Sturm-Liouville
spectral problem [5,6]
$$P_{v+1}(\lambda) = [\lambda - \Delta (v + l_0)] P_{v}(\lambda) -
|g|^{2} \psi (l_0+v; [l_i]) P_{v-1}(\lambda), \,
v=0,\cdots,s;    \eqno (3.7a)$$
$$P_{0}(\lambda) =1,\quad P_{-1}(\lambda) = 0 = [\lambda - \Delta (s +
l_0)] P_{s}(\lambda) - |g|^{2} \psi (l_0+s; [l_i]) P_{s-1}(\lambda) 
\eqno (3.7b)$$
for finding non-classical orthogonal (in view of (3.5b)) polynomials
$$P_{v}(\lambda) = (g^* )^v N^{-2}(v;[l_i])\tilde Q_{v}([l_i];\lambda)/
\tilde Q_{0}([l_i];\lambda)= (g^* )^v N^{-1}(v;[l_i])
Q_{v}([l_i];\lambda)/ Q_{0}([l_i];\lambda)
  \eqno (3.8)$$
for finding  non-classical orthogonal (in view of Eqs. (3.5b) polynomials
in the discrete variable $\lambda$ on the non-uniform lattice
$\{\lambda_f([l_i])\}_{v=0}^s$ [5]. Indeed, Eqs. (3.7), (3.8) provide
the following (easily realized by means of FORTRAN programs) algorithm
for solving the eigenproblem (3.2).\\

 i) Using the recursive formula (3.7a)  with initial values from Eq.
(3.7b)
 one calculates the  polynomial sequence $\{P_{v}(\lambda)\}_{v=0}^s$.

 ii) Inserting $P_{s-1}(\lambda), P_{s}(\lambda)$ in the last equality in
 (3.7b) one gets the algebraic equation with respect to $\lambda$; its
 solution yield the  sequence $\{\lambda_f([l_i])\equiv
\lambda_f\}_{f=0}^s$
 of admissible values of the spectral parameter $\lambda$ and the
appropriate energy spectrum $\{E_f = E_{f}([l_i])\}_{f=0}^s$.

 iii)  For each value $\lambda_{f}([l_i])$ using
 $\{P_{v}(\lambda)\}_{v=0}^s$ and Eq. (3.8) one finds the sequence
 $\{Q_{v}^{f}([l_i])= Q_{v}([l_i];\lambda_f)\}_{v=1}^s$ of all amplitudes
as functions in the only undetermined quantity $Q_0(\lambda_f)=Q_{0}([l_i];
 \lambda_f)$ which, in turn, is found from the normalization condition of
 Eqs. (3.5b).

In order to make numerical calculations we implemented this
algorithm with the help of the REDUCE Package [14] similar to the
 algebraic construction in [17].

\section{Numerical analysis for the second-harmonic generation model}

In order to examine the efficiency of calculation schemes and the
algorithm given above  we tested them  by means of computer
experiments for the resonance second-harmonic-generation model
widely examined in quantum optics [1,8,10] and determined by
$H^{hg}(1;2)$ with $\omega_0=2\omega_1$. In this case, according
to Eqs. (2.6b), (2.8a), (2.9a), and (3.3b) we have
$$\psi (l_0+v+1; [l_i])= (k + 2v+2)(k + 1  + 2v)(s - v),\; k=0,1; s=
0,1,\dots, \eqno (4.1a)$$
$$ C(l_1=\frac{k+2s}{3})=\omega_1(k+2s),\; \Delta=0,\;
\lambda_v^{cmf}([l_i];
r) =  - 2|g| (j - v)\sin 2r\,\sqrt{\phi [(- j+v)\cos 2r]},$$
$$\phi (- j+f)= 2[s+2k+1- (s-2f)],\;j= s/2
\eqno (4.1b) $$

We implemented calculations of exact  values $\lambda_f([k,s])$ according
to the algorithm of Section 3 and of their approximations $\lambda_f^{cmf}
([k,s];r_i)$ according to  Eq. (4.1b) for $ g=1,  k=0,1, s=20,100,500,1000,
10000$. Values of the fitting parameter  $r$ were determined from
energy-stationarity-conditions: $2r_1= \arccos \frac{1}{3}$ (optimizing
only the upper part of spectra) [8], $2r_3=\arccos 0=\frac{\pi}{2}$
(quasi-linear approximation) [5,8] and from minimizing the proximity measure 
(3.4a): $2r_2= \arccos\frac{1}{\sqrt{s}}$
("smooth" cluster mean-field approximation) [8]; herewith $\lambda_f^{cmf}
([k,s];\mp r_1)$ means that we take  $r=-r_1$ in the first half of spectra
and  $r=r_1$ the second one. To estimate the accuracy
of approximations we used, besides Eq. (3.4a) the measures
$$\delta^2_E([l_i]) =\frac{\sum_{v=0}^s [(\lambda_v ([l_i]) -
\lambda_v^{cmf}([l_i];r)]^2} {\sum_{v=0}^s (\lambda_v ([l_i]))^2},\quad
\delta^2_{E_{up}}([l_i]) =\frac{\sum_{v=s/2}^s [(\lambda_v ([l_i]) -
\lambda_v^{cmf}([l_i];r)]^2} {\sum_{v=s/2}^s (\lambda_v ([l_i]))^2}
 \eqno (4.2)$$
which characterize energy spectra more precisely in comparison with Eq.
(3.4a).

Typical results of numerical calculations obtained are presented in two
tables.\\

\centerline{Table 1. Multiplets
$\{\lambda_f([0,s])\},\{\lambda_f^{cmf} ([0,s];r_i)\}$ as
functions in  $f=0,1,...,s$ for $s=10^2$}

\begin{tabbing}
  v \qquad\quad\= $\lambda_f([0,s])$\qquad\=$\lambda_f^{cmf}([0,s];r_1)$
\quad\=$\lambda_f^{cmf}([0,s];\mp r_1)$\quad\=$\lambda_f^{cmf}([0,s];
r_2)$ \quad\=$\lambda_f^{cmf}([0,s];r_3)$ \\
  0 \>      -1536.9\>  -1096.7\>  -1545.3\>  -1482.4\>  -1421.2\\
 10 \>      -1151.7\>   -919.6\>  -1205.2\>  -1175.2\>  -1137.0\\
 20 \>       -798.1\>   -720.0\>   -880.0\>   -873.3\>   -852.7\\
 30 \>       -480.3\>   -499.3\>   -570.2\>   -576.7\>   -568.5\\
 40 \>       -205.5\>   -259.0\>   -276.7\>   -285.6\>   -284.2\\
 50 \>          0.0\>      0.0\>      0.0\>      0.0\>      0.0\\
 60 \>        205.5\>    276.7\>    276.7\>    280.0\>    284.2\\
 70 \>        480.3\>    570.2\>    570.2\>    554.3\>    568.5\\
 80 \>        798.1\>    880.0\>    880.0\>    822.8\>    852.7\\
 90 \>       1151.7\>   1205.2\>   1205.2\>   1085.5\>   1137.0\\
100 \>       1536.9\>   1545.3\>   1545.3\>   1342.3\>   1421.2\\
$\delta^2_H=$     \> \>   10.222\>  -12.220\>    0.010\>   -1.000\\
$\delta^2_E=$     \> \>    2.563\>    0.670\>    0.806\>    0.657\\
$\delta^2_{E_up}=$\> \>    0.670\>    0.670\>    0.944\>    0.657\\
\end{tabbing}

\centerline{Table 2. Multiplets  $\{\lambda_f([0,s])\},\{\lambda_f^{cmf}
([0,s];r_i)\}$ as functions in  $f=0,...,s$ for $s= 10^4$}

\begin{tabbing}
  v \qquad\quad\= $\lambda_f([0,s])$\qquad\=$\lambda_f^{cmf}([0,s];r_1)$
\quad\=$\lambda_f^{cmf}([0,s];\mp r_1)$\quad\=$\lambda_f^{cmf}([0,s];
r_2)$ \quad\=$\lambda_f^{cmf}([0,s];r_3)$ \\
    0\>  -1539573\> -1088743\> -1539658\> -1421266\> -1414284\\
  200\>  -1460154\> -1055592\> -1470663\> -1364145\> -1357712\\
  600\>  -1304855\>  -986414\> -1334422\> -1249971\> -1244570\\
 1000\>  -1154419\>  -913501\> -1200541\> -1135886\> -1131427\\
 2000\>   -800913\>  -715586\>  -876392\>  -851069\>  -848570\\
 3000\>   -483083\>  -496534\>  -567801\>  -566815\>  -565713\\
 4000\>   -208106\>  -257638\>  -275425\>  -283125\>  -282856\\
 4400\>   -113484\>  -156775\>  -163176\>  -169807\>  -169714\\
 4800\>    -31728\>   -52979\>   -53690\>   -56579\>   -56571\\
 5000\>         0\>        0\>        0\>        0\>        0\\
 5200\>     31728\>    53690\>    53690\>    56557\>    56571\\
 5600\>    113484\>   163176\>   163176\>   169603\>   169714\\
 6000\>    208106\>   275425\>   275425\>   282559\>   282856\\
 7000\>    483083\>   567801\>   567801\>   564553\>   565713\\
 8000\>    800913\>   876392\>   876392\>   845978\>   848570\\
 9000\>   1154419\>  1200541\>  1200541\>  1126836\>  1131427\\
 9400\>   1304855\>  1334422\>  1334422\>  1239020\>  1244570\\
 9800\>   1460154\>  1470663\>  1470663\>  1351113\>  1357712\\
10000\>   1539573\>  1539658\>  1539658\>  1407125\>  1414284\\
$\delta^2_H=$     \> \>   11.102\>  -11.122\>    0.000\>   -0.010\\
$\delta^2_E=$     \> \>    2.577\>    0.590\>    0.626\>    0.625\\
$\delta^2_{E_up}=$\> \>    0.590\>    0.590\>    0.639\>    0.625\\
\end{tabbing}

As is seen from data given  in Tables 1, 2, there
is  a satisfactory consent of exact results and those obtained
with the help of the approximate formula (3.3b)  at $s \gg 1$ in
the most parts of energy spectra for $r=r_{i=2,3}$ and $r=\mp
r_1$.  Discrepancies between exact and approximate results in
middle parts of spectra are explained by the availability of the
square-root singularities  in $H^{cmf}(\xi)= \sum_{v,[l_i]}\,
E^{qc}_v([l_i];\xi) |E^{cmf}_v([l_i];\xi)\rangle \langle
E^{cmf}_v([l_i];\xi)|$, and the symmetry breaking of spectra at
$r=r_{i=1,2}$  is due to smoothing this singularities within the
framework of the "smooth" $su(2)$ - quasiclassical approximation
[8,9]. A more full analysis of results of computer experiments
above  will be given elsewhere.

 \section{Conclusion}
So, we developed an algebraic scheme and routines (implemented
with the help of the REDUCE Package [14]) to get exact solutions
of both spectral and evolution problems for a wide class of
integrable models of quantum optics having PLA  $su_{pd}(2)$ as
their  dynamic symmetry algebras. These results along with
appropriate computer realizations of Eqs. (3.3) provide an
efficient tool for examining models under study in appropriate
(determined by computer powers) ranges of characteristic model
parameters given by $l_i$. Numerical calculations given in
Section 4 showed a good consent of exact and  approximate results
at $s \gg 1$ and at relevant choices of the fitting parameter $r$
in (3.3b). Furthermore, the approximate  solutions of the
eigenproblem (3.2) can be improved with the help of the algebraic
perturbation algorithm like that developed in [17] for the
hydrogen atom or by means of using  modifications of algebraic
schemes of the work [6]. Therefore, the exact  calculation
schemes developed above, when being completed by quasiclassical
calculations [8] based on formulas like Eq. (3.3) as well as
their improvements, can be used for analyzing models  under
consideration in all ranges of characteristic parameters $l_i$.
The work along these lines is in progress.
\newcommand{\etal}{{\em et al.}}
\setlength{\parindent}{0mm}
\vspace{5mm}
{\bf References}
\begin{list}{}{\setlength{\topsep}{0mm}\setlength{\itemsep}{0mm}%
\setlength{\parsep}{0mm}}
\item[1.] J. Perina, Quantum Statistics of Linear and
Nonlinear Optical Phenomena. (Reidel, Dordrecht  1984).
\item[2.] J.H. Eberly, N.B. Narozhny, J.J. Sanchez-Mondragon,
Phys. Rev. Lett., {\bf 44}, 1329 (1980).
\item[3.] V.P. Karassiov, L.A. Shelepin, Trudy FIAN [P.N. Lebedev
Institute Proc.], {\bf 144}, 124 (1984).
\item[4.]  V.P. Karassiov, J. Sov. Laser Res., {\bf 12}, 147 (1991).
\item[5.] V.P. Karassiov, Teor. Mat. Fiz., {\bf 95}, 3 (1993); J. Phys.,
{\bf A 27}, 153 (1994).
\item[6.] V.P. Karassiov and A.B. Klimov, Phys. Lett., {\bf A 189}, 43
(1994).
\item[7.] V.P. Karassiov, Rep. Math. Phys., {\bf  40},235 (1997); Czech.
J. Phys.,{\bf 48}, 1381 (1998).
\item[8.] V.P. Karassiov,  Phys.  Lett., {\bf A 238}, 19 (1998); J. Rus.
Laser Res., {\bf 20}, 239 (1999).
\item[9.] V.P. Karassiov, J. Rus. Laser Res., {\bf 21}, 370 (2000).
\item[10.] S.M. Chumakov and Kozierowski, Quantum Semiclass. Opt. {\bf 8},
775
(1996); A. Bandilla, G. Drobny and I. Jex, Phys. Rev., {\bf A 53} 507
(1996).
\item[11.] A.M.  Perelomov, Generalized Coherent States and Their
Applications. (Nauka, Moscow, 1987).
\item[12.] L.A. Takhtajan L. and  L.D. Faddeev, Uspekhi Math. Nauk
{\bf 34}, N 5, 13 (1979); B. Jurco, J. Math. Phys. {\bf 30}, 1289
(1989).
\item[13.] M. Rocek,  Phys. Lett., {\bf  B 255}, 554 (1991).
\item[14.] A.S Hearn. Reduce User's Manual. Ver. 3.6.
Santa-Monica, CA, USA. 1995.
\item[15.] D.A. Varshalovich, A.N. Moskalev, V.K. Khersonsky.
Quantum Theory of Angular Momentum.(Nauka, Leningrad, 1975).
\item[16.] L.G. Jaffe, Rev. Mod. Phys., {\bf 54} 407 (1982); A.
Chatterjee, Phys. Rep., {\bf 186}, 249 (1990).
\item[17.] A. Gusev, V. Samoilov, V. Rostovtsev and S. Vinitsky, in:
Computing Algebra in Scientific Computing: Proc. Workshop  CASC
2000, V.G. Ganzha, E.W. Mayr, E.V. Vorozhtsov (Eds.) (Springer,
Berlin e.a., 2000), p. 219.
\end{list}
\end{document}